\documentclass{article}
\usepackage{spconf,amsmath,graphicx}
\usepackage{multicol}
\graphicspath{{images/}}

\title{Automatic Instrument Recognition in Polyphonic Music Using Convolutional Neural Networks}
\name{Peter Li, Jiyuan Qian, and Tian Wang\sthanks{All authors contributed equally to this work}}
\address{New York University \\ Center for Data Science \\ New York, NY 10003}
\begin{document}
%
\maketitle
\begin{abstract}
Traditional methods to tackle many music information retrieval tasks typically follow a two-step architecture: feature engineering  followed by a simple learning algorithm. In these "shallow" architectures, feature engineering and learning are typically disjoint and unrelated. Additionally, feature engineering is difficult, and typically depends on extensive domain expertise. 

In this paper, we present an application of convolutional neural networks for the task of automatic musical instrument identification. In this model, feature extraction and learning algorithms are trained together in an end-to-end fashion. We show that a convolutional neural network trained on raw audio can achieve performance surpassing traditional methods that rely on hand-crafted features. 
\end{abstract}

\section{Introduction}

Computer audition is the general study of the systems and methods necessary for audio understanding by a machine. In a sense, computer audition concerns itself with the study of designing computers that can ``hear'' as humans do. The goal is a machine that can ``organize what they hear; learn names for recognizable objects, actions, events, places, musical styles, instruments, and speakers; and retrieve sounds by reference to those names." \cite{lyon}

In this paper, we focus on the first two tasks. Given a musical recording, how can we train a system to identify the instruments that are present? We present an application of deep learning for the task of automatic musical instrument identification in polyphonic music. We show that an end-to-end system using convolutional neural networks trained on raw audio can surpasses traditional MIR models trained using hand-crafted features.

\subsection{Relation to Previous Work}

In a paper calling for the adoption of deep architectures in MIR,\cite{da} describe traditional MIR methods as follows:

\begin{quote}

The traditional approaches to these problems are rather homogeneous, adopting a two-stage architecture of feature extraction and semantic interpretation, e.g. classification, regression, clustering, similarity ranking, etc. Feature representations are predominantly hand-crafted,drawing upon significant domain-knowledge from music theory or psychoacoustics. 

\end{quote}

Since good features are hard to craft, much of the recent research in the MIR community has concerned itself with the semantic interpretation part of the problem i.e. training better models given a set of standard audio features (e.g. Mel-Frequency Cepstral Coefficients or chroma)\cite{da}. 

In this paper, we depart from the traditional MIR approach. Using convolutional neural networks, we train a model using raw audio as input. We utilize a deep architecture where feature extraction \emph{and} semantic interpretation can both be learned from data directly. 

This approach to audio signal processing has been explored before in speech processing e.g., \cite{ron} and musical audio tagging \cite{sander}. However, to the best of our knowledge, this is the first application of deep learning to source identification. 

\begin{figure*}[!ht]
\label{cnnLayer}
  \begin{tabular}{@{} c@{ } c@{ } c@{ } c@{} @{ } @{ } c@{ } c@{ } c@{ } c@{ }}
    \includegraphics[height=0.1\linewidth]{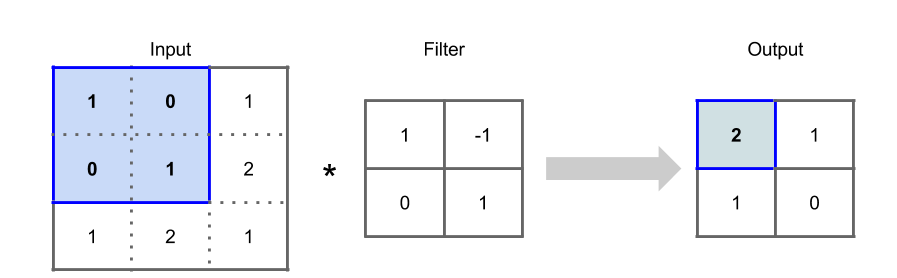}&
    \includegraphics[height=0.1\linewidth]{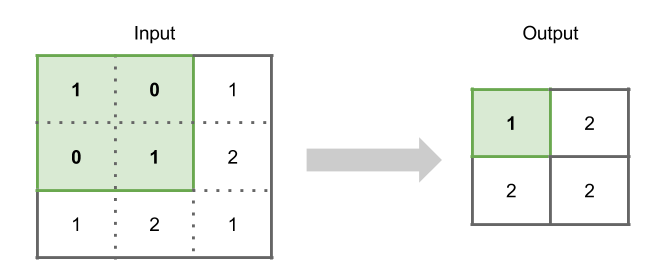}&
    \includegraphics[height=0.1\linewidth]{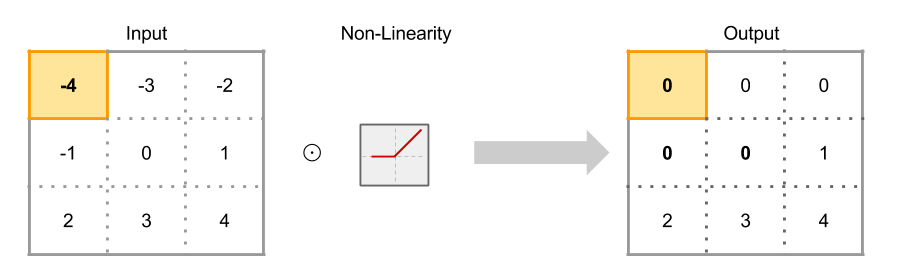}\\[-0.8em] \\
    a) Convolution & b) Pooling & c) Non-Linearity\\[0.8em]
  \end{tabular}
  \caption{Common ConvNet Layers}
\end{figure*}

\begin{figure*}[!ht]
\label{cnnArch}
    \begin{center}
		\includegraphics[scale=0.3]{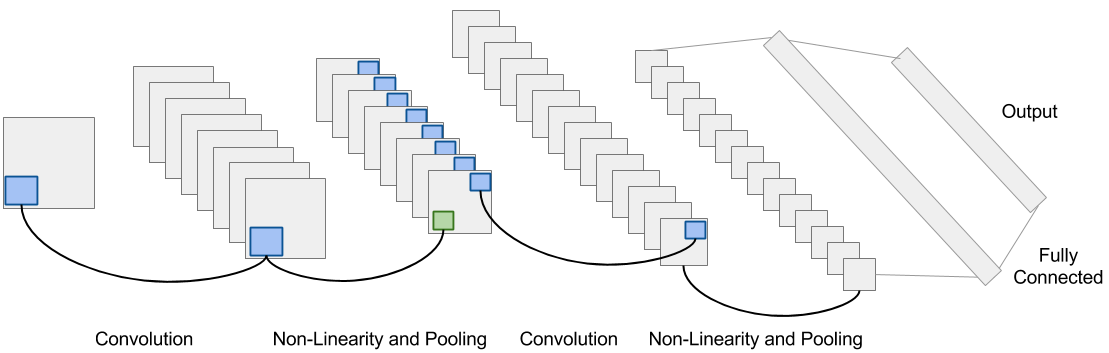}
		\caption{Sample ConvNet Architecture}
	\end{center}
\end{figure*}

\section{Problem Definition}

Before going into the details of our work, we give a formal definition of our task: \\
\\
Given a section of audio $x$, we would like to predict a vector $y \in \{0,1\}^l$ where $l$ is the total number of instruments and 

  \begin{equation*}
    y_{i}=
    \begin{cases}
      1, & \text{if}\ \text{instrument} \ i \text{ is in } x \\
      0, & \text{otherwise}
    \end{cases}
  \end{equation*}
  
We treat this as a multi-label classification problem where each label corresponds to an instrument. A model should output 1 if an instrument is present and 0 otherwise.

\section{Model}

Convolutional Neural Networks (CNN) \cite{ConvNet} can be seen as a trainable feature extractor coupled with a learning model. Generally, they are know to be good at extracting high level features that represent abstract concepts from the original data. A typical model contains multiple layers of modules, where each module performs a simple data transformation. In Figure 1, we show a few common operations on a general matrix input. In a convolution layer, a filter, whose weights are learned, is convolved with its input by taking point-wise multiplication and then summing the results. Pooling is a down-sampling operation that combines nearby points. Non-linearities are applied point wise. In Figure 2, we show a sample model architecture.  Each of the first two layers contains a convolution, pooling, and non-linearity operation. The final two layers are a fully connected neural network. 

Hyper-parameters of the system include the number of layers and feature maps, convolution filter size, pooling size and which non-linear activation function to use. The choice of activation functions usually include the sigmoid function, hyperbolic tangent function and more recently Rectified Linear Unit (ReLU)\cite{dropout}. \\

\noindent\textbf{Model Architecture}\\

Our convolutional neural network contains three temporal convolutional layers (a convolution operation where the height of the filter is the height of the input) with ReLU and max pooling. These three layers are followed by two fully connected layers with ReLU and Dropout on the first layer and a sigmoid after the the second fully connected layer.  This gives us a $11 \times 1$ vector $\hat{y}$ where each $\hat{y_i}\in[0,1]$. These predicted activations are compared to training activations using a binary cross entropy loss function. The exact model Specifications are presented in Table 1.

\begin{table}[h]
\label{convnet_spec}
\begin{center}
\begin{tabular}{llll}
\multicolumn{1}{c}{\bf Parameter}  &\multicolumn{1}{c}{\bf }
\\ \hline \\
\verb!Convolution Feature Maps 1!            &256\\
\verb!Convolution Filter Size 1!         &3101\\
\verb!Maxpooling Stride Size 1!            &20\\
\verb!Maxpooling Size 1!            &40\\
\verb!Convolution Feature Maps 2!            &384\\
\verb!Convolution Filter Size 2!         &300\\
\verb!Maxpooling Size 2!            &30\\
\verb!Maxpooling Stride Size 2!            &20\\
\verb!Convolution Feature Maps 3!            &384\\
\verb!Convolution Filter Size 3!         &20\\
\verb!Maxpooling Size 3!            &8\\
\verb!Maxpooling Stride Size 3!            &4\\
\verb!Layer 4 Output Size!             &400\\
\verb!Final Output Size!           &11\\
\end{tabular}
\end{center}
\caption{ConvNet Specifications}
\end{table}

\section{Experiments}

\subsection{Data and Setup}
We trained and evaluated our model using data from MedleDB \cite{MedleyDB}. MedleyDB is a multitrack dataset of 122 annotated musical recordings. For each song we have three types of audio content: mix, stems, and raw audio. Mixed audio is comprised of a set of stems and stems are comprised of a mix of raw audio. Since MedleyDB was primarily created to support research on melody extraction, we had to create our own labels for model training. In the follow sections, we outline our procedure. 

For each stem, we have annotations on instrument activation that represent the confidence of whether or not the instrument is active during that time frame. These annotations were generated using a standard envelope following technique on each stem, consisting of half-wave rectification, compression, smoothing, and down-sampling. For additional details see \cite{MedleyDB}.\\

\noindent\textbf{Dataset Split} \\

To train our model, we sliced individual tracks into one second, non-overlapping clips. 80\% of the clips were used for model training and the remaining 20\% were reserved for testing. When splitting the data, there were two main considerations. First, we didn't want to have labels that were in the test set but not in the training set. Second, we didn't want to have clips belonging to the same track in both the test and training sets. 

To address both issues at the same time, we use the algorithm in \cite{datasplit} to split the 122 mixed tracks into a training and test set based on the instruments that appear in each track. After the initial split, we cut each track into one second non-overlapping clips. This gives us our final training and test sets. Because tracks vary in length, after cutting into one second clips, the training and test set have 21177 and 4985 clips respectively.\\

\noindent \textbf{Label Generation}\\

To train our model, we need labels indicating whether or not an instrument appears in the entire audio clip. However, the activation confidence scores from MedleyDB provide local information i.e, activations at each point in time. To convert this to a global label for the entire clip, we take the maximum of the moving average of the activation confidences. An instrument is considered active in a clip as long as its average activation confidence exceeds a threshold over a particular window. For our experiment, we chose a windows length of 100ms and a threshold of 0.5. Figure  3 shows an example of this procedure. 

\begin{figure}[htbp]
\label{activation}
\begin{center}
\includegraphics[height=0.5\linewidth]{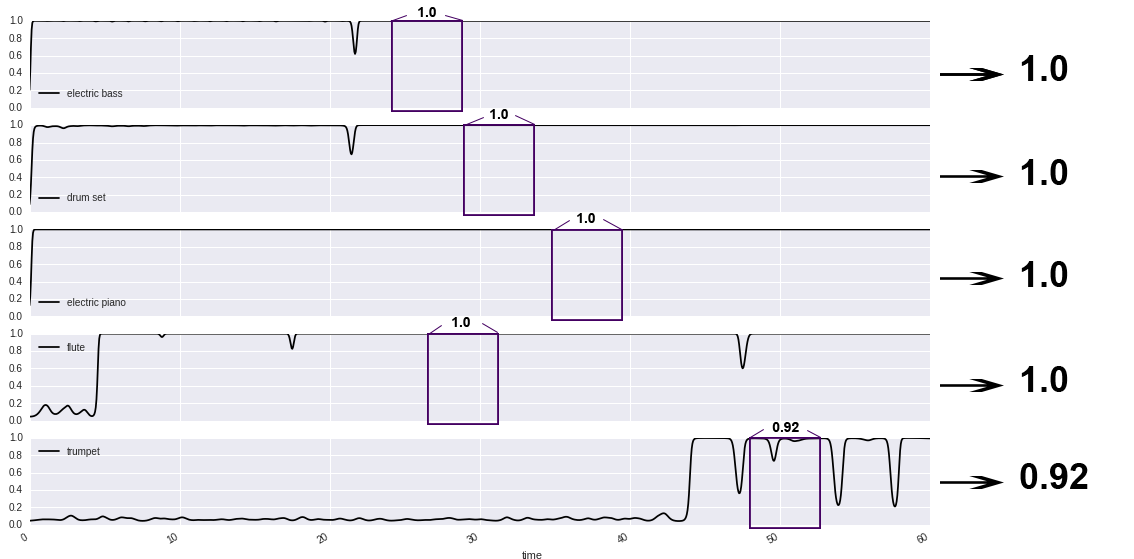}\\
\caption{Label Generation Procedure}

\end{center}
\end{figure}

The activation annotations cover 82 instrument. We grouped these instruments into 70 categories. To simplify model training and to have a more balanced dataset for learning, we combined all categories appearing in less than 20 songs into an `OTHER' category. This gives us 11 classes for classification:

    \vspace{-0.3cm}
    
    \begin{multicols}{2}
    		\noindent $\bullet$ electric bass\\
    		$\bullet$ acoustic guitar\\	
    		$\bullet$ synthesizer\\  		
    		$\bullet$ drum set\\
    		$\bullet$ fx/processed sound\\  			
    		\columnbreak
    		$\bullet$ voice\\
    		$\bullet$ violin\\
    		$\bullet$ piano\\
    		$\bullet$ distorted electric guitar\\
    		$\bullet$ clean electric guitar\\
    		$\bullet$ OTHER
	\end{multicols}

\subsection{CNN Training}

\begin{table*}[ht]
\label{exp_res}
\begin{center}
\begin{tabular}{lllllll}
\multicolumn{1}{c}{\bf Models}  &\multicolumn{1}{c}{\bf Accuracy} &\multicolumn{1}{c}{\bf Exact Match} &\multicolumn{1}{c}{\bf Precision} &\multicolumn{1}{c}{\bf Recall} &\multicolumn{1}{c}{\bf F-micro} &\multicolumn{1}{c}{\bf F -macro}
\\ \hline \\
\verb!Audio + CNN!         &\textbf{82.74}\%&\textbf{25.78}\%&0.7560&\textbf{0.6888}&\textbf{0.7208}&\textbf{0.6433} \\
\verb!MFCC + Random Forest!      &82.13\%&17.53\%&\textbf{0.7908}&0.5400&0.6418&0.4471 \\
\verb!MFCC + Logistic Regression!      &81.80\%&18.17\%&0.7457&0.5857&0.6561&0.4840 \\
\verb!Predict Majority Class!     &70.37\%&9.95\%&0.5001&0.4602&0.4793&0.1801 \\
\end{tabular}
\end{center}
\caption{Experiment Results}
\end{table*}

Our CNN was trained using stochastic gradient descent with a batch size of 16. To speed up training time, we transformed inputs using global contrast normalization. This is a standard preprocessing step and has been show to speed up training time. \cite{efficientbp}. 

\subsection{Benchmarks}

For comparison, we also ran tests using more traditional MIR techniques. In the benchmarks, we use domain knowledge to construct music related features. For each audio clip, we first computed the Mel-frequency cepstral coefficients (MFCC) along with the first and second order differences, MFCC $\Delta$ and $\Delta^2$. These three matricies where stacked and modelled using a Gaussian distribution \cite{Logan}.

\begin{equation*}
\quad
\begin{pmatrix}
MFCC\\
MFCC\Delta\\
MFCC\Delta^2
\end{pmatrix}
\quad = \quad\begin{pmatrix}
\vert &  & \vert \\
m_1 & \hdots & m_T \\
\vert &  & \vert \\
\end{pmatrix}
\quad,
\end{equation*}
where $m_i \sim N(\mu, \Sigma)$

$(\mu, \Sigma)$ were used as features to train a random forest and logistic regression.

\subsection{Experiment Results}

The results of our experiments are shown in Table 2.  In addition to the random forest and logistic regression, we also provide a naive benchmark that always predicts the three most common instruments in the training set. As seen, the Audio + CNN model generally outperforms the baseline models. 

\section{Discussion}

Convolutional neural networks have recently show remarkable results in a number of tasks. However, it is often times difficult to intuitively understand what they are doing and why they work. In this section, we make an attempt at examining the weights learned in the first convolutional layer. This sections follows the procedures of previous works in training CNN's on audio waveforms and confirms previous findings. 

Examining the filter weights learned in the first layer, we see that the model does seem to learn a set of frequency selective filters. In Figure 4, we plot the magnitude spectra of each filter, sorted by dominant frequency. Similar to results in \cite{sander} and \cite{ron}, the first convolutional layer seems to learn an auditory scale filter bank.

\begin{figure}[h]
\begin{center}
\includegraphics[width=\columnwidth]{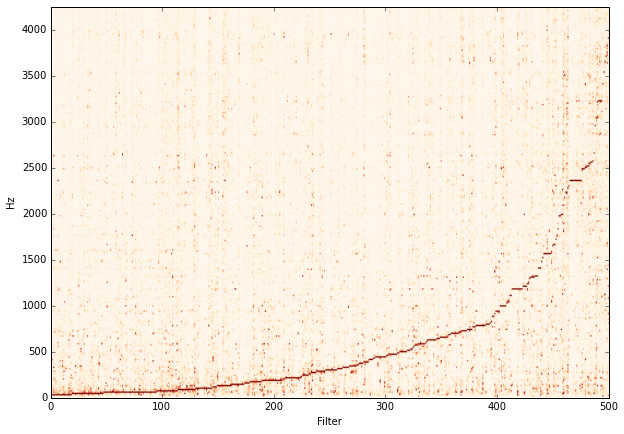}\\
\caption{Rescaled magnitude spectra sorted by dominant frequency. The spectra of each filter was rescaled to [0,1] by subtracting the minimum and dividing by the range. }
\end{center}
\end{figure}

In Figure 5, we show a sample of filters learned in the first layer. As in \cite{sander} many of the learned filters are translated versions of each other. This is not necessary surprising since phase invariance is most likely difficult learn given our architecture. 

\begin{figure}[h]
\label{activation}
\begin{center}
\includegraphics[width=\columnwidth]{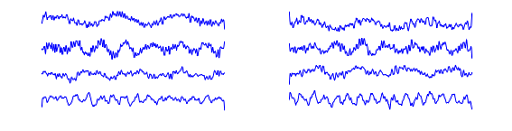}\\
\caption{Sample of filters learned in the first layer.Filters were low pass filtered to remove noise.}
\end{center}
\end{figure}

\section{Conclusion and Future Work}

We present a convolutional neural network for instrument identification. We show that an end-to-end deep learning system can be trained to achieve performance in line with (and sometimes exceeding) traditional methods that rely on extensive domain knowledge.

In future work, we will investigate methods to further understand the transformations made by the CNN. Additionally, we would like to explore different architecture that may be able to learn phase and translation invariance. 

\subsubsection*{Code}

Code for this project can be found on \\ https://github.com/glennq/instrument-recognition.

\vfill
\pagebreak


\bibliographystyle{IEEEbib}
\clearpage
\bibliography{strings,refs}

\end{document}